\documentclass[9pt,twocolumn,twoside]{pnas-new}

\templatetype{pnasbriefreport} %

\setboolean{displaywatermark}{false}
\usepackage{academicons}
\definecolor{orcidlogocol}{HTML}{A6CE39}
\usepackage{xcolor}
\newcommand{\orcid}[1]{\href{https://orcid.org/#1}{\textcolor[HTML]{A6CE39}{\aiOrcid}}}

\title{Cities in a world of diminishing transport costs} 
\author[a,b,1]{Tomoya Mori}
\author[a]{Minoru Osawa} 

\affil[a]{Institute of Economic Research, Kyoto University. Yoshida-Honmachi, Sakyo-Ku, Kyoto, Kyoto 606-8501, Japan.}
\affil[b]{Research Institute of Economy, Trade and Industry, 11th floor, Annex, Ministry of Economy, Trade and Industry 1-3-1, Kasumigaseki Chiyoda-ku, Tokyo 100-8901, Japan.}

\leadauthor{Mori} 

\authorcontributions{Author contributions: T.M. and M.O. equally contributed to the design and implementation of the research, to the analysis of the results, and to the writing of the manuscript.}
\authordeclaration{The authors declare no conflict of interest.}
\correspondingauthor{\textsuperscript{1}To whom correspondence should be addressed. E-mail: mori@\@kier.kyoto-u.ac.jp}

\keywords{city size $|$ spatial pattern $|$ spatial scale $|$ science of cities} 

\begin{abstract}
Economic activities favor mutual geographical proximity and concentrate spatially to form cities. 
In a world of diminishing transport costs, however, the advantage of physical proximity is fading, and the role of cities in the economy may be declining. 
To provide insights into the long-run evolution of cities, 
we analyzed Japan's census data over the 1970--2015 period. 
We found that fewer and larger cities thrived at the national scale, suggesting an eventual mono-centric economy with a single megacity; 
simultaneously, each larger city flattened out at the local scale, suggesting an eventual extinction of cities. 
We interpret this multi-scale phenomenon as an instance of pattern formation by self-organization, which is widely studied in mathematics and biology. 
However, cities' dynamics are distinct from mathematical or biological mechanisms because they are governed by economic interactions mediated by transport costs between locations.  
Our results call for the synthesis of knowledge in mathematics, biology, and economics to open the door for a general pattern formation theory that is applicable to socioeconomic phenomena. 
\end{abstract}

\dates{This manuscript was compiled on \today}
\doi{\url{www.pnas.org/cgi/doi/10.1073/pnas.XXXXXXXXXX}}

\begin{document}

\maketitle
\thispagestyle{firststyle}
\ifthenelse{\boolean{shortarticle}}{\ifthenelse{\boolean{singlecolumn}}{\abscontentformatted}{\abscontent}}{}

\dropcap{C}ities are home to most of the population and economic activities within developed countries and industrialized regions in developing countries \cite{UN2018}. 
For instance, Japanese cities accounted for 80\% of the total national population, but occupied only 6\% of the total land area in Japan in 2015 according to our data. 
Agglomeration effects, or superlinear scaling in cities' outputs \cite{Bettencourt-et-al-2007}, facilitate city existence and growth. 
The fundamental trade-off between the various agglomeration effects and the costs associated with the transfer of goods, people, and information has been the key factor for understanding the spatial organization of economic activities \cite{Fujita-Thisse-Book1996}. 
The advantage of physical proximity and the role of cities in the economy might diminish in the near future, given the continued improvement in transport technology and the rise of Internet communication in recent decades \cite{Cairncross-Book1997}.
Instead, cities might continue to thrive even in the era of diminishing transport costs because in-person exchange of ideas is at the heart of innovation \cite{Glaeser-Book2011}. 

To gain insights into the fate of cities, we address the actual situation of cities in Japan between 1970 and 2015. 
Japanese data in this period have unique and ideal features for studying the impacts of transport costs on the population distribution. 
First, Japan experienced from-scratch development in its nationwide high-speed transport networks, which highlights the role of decline in transport costs in a relatively short period of time compared to the situation in other developed economies. 
Second, the availability of high-geographic-resolution population data over the study period enables us to assemble a new bottom-up dataset of cities, which, in turn, allows us to evaluate the evolution of the spatial population distribution in detail without being affected by the municipal boundaries.

\section*{Results}

\begin{figure}
\centering
\includegraphics[height=\vsize]{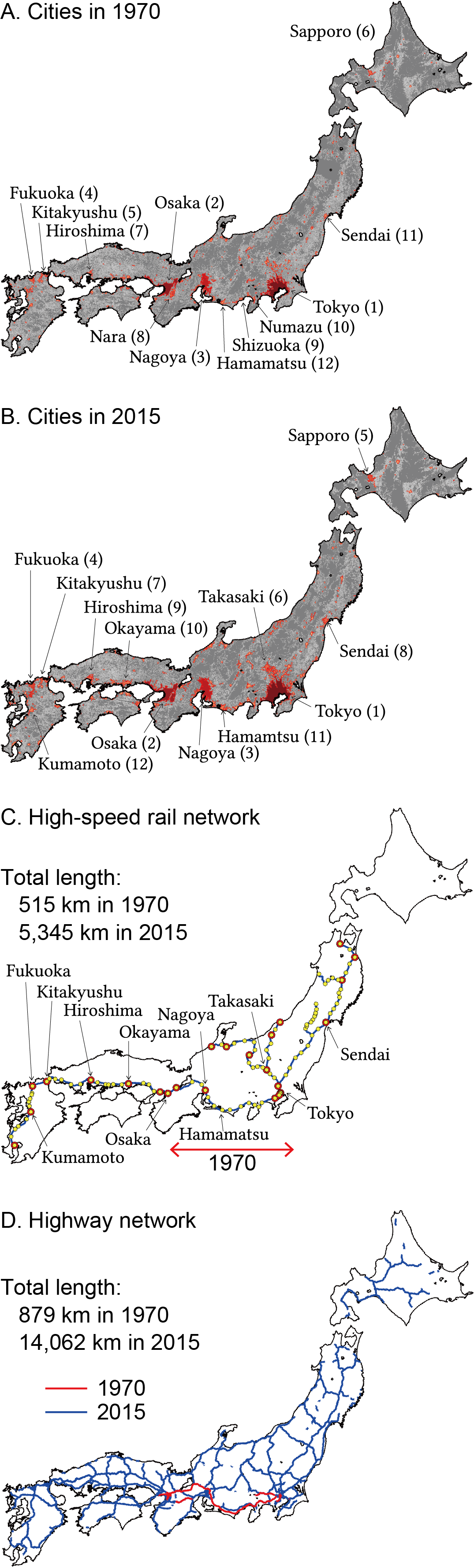}
 \caption{Spatial patterns of the population and transport network in Japan. 
 (A, B) The red areas on each map indicate the 503 and 450 cities in 1970 and 2015, respectively. Otherwise, a lighter color corresponds to a larger population size per a 1km-by-1km grid. We consider four major islands (Honshu, Kyushu, Shikoku, and Hokkaido) as well as other islands connected to one of these by road. The largest 12 cities are indicated with their population rankings between parentheses. 
 (C) The high-speed railway network in 2015 is indicated by the blue lines. The yellow circles indicate the locations of stations and the red circles the stations at which express trains stop. In 1970, only the segment between Tokyo and Osaka (the red arrow) was completed. 
 (D) The highway network as of 1970 (red) and 2015 (blue).  
 Highway and high-speed railway network data are obtained from Kokudo Suuchi Joho Download (\url{https://nlftp.mlit.go.jp/ksj/index.html}). 
 \label{fig:ua}}
\end{figure}

We identify Japanese cities in a bottom-up manner based on grid-level population census data (see \textit{Materials and Methods}).  
Figs.~\ref{fig:ua}A and B show the 503 and 450 cities respectively identified in 1970 and 2015. 
Japan experienced significant decline in interregional transport costs in this period. 
Triggered by the Tokyo Olympics of 1964, the total high-speed railway (highway) length in Japan increased from 515 km (879 km) in 1970 to 5,345 km (14,062 km) in 2015.  
In 1970, the high-speed railway and highway networks connected only Tokyo and Osaka, the two largest cities (Figs.~\ref{fig:ua}C and D); 
four of the ten largest cities were located between Tokyo and Osaka in 1970 (Fig.~\ref{fig:ua}A).
However, in 2015, the largest cities were geographically farther apart, as only one among the ten largest cities (Nagoya) remained between Tokyo and Osaka  (Fig.~\ref{fig:ua}B). 
\begin{figure*}
\centering{}
\includegraphics[width=\hsize]{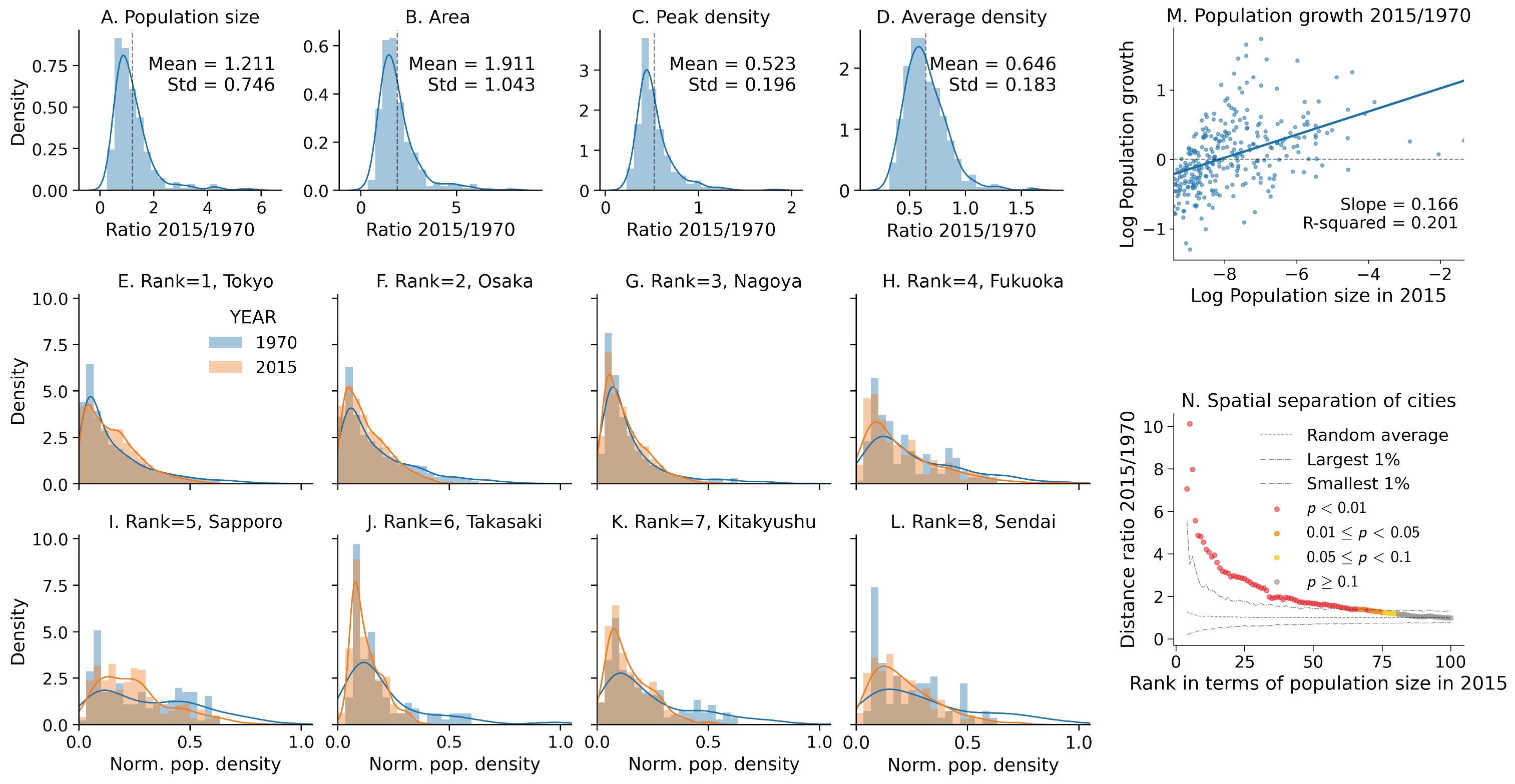}
 \caption{(A--D) Population size, area, largest peak population density, average population density for the 302 Japanese cities existing both in 2015 and 1970. In graphs A, C, and D, the total national population is normalized be 1 in each year. (E--L) The distribution of population size in a 1km-by-1km cell (i.e., population density) in 1970 and 2015 in each of the largest eight Japanese cities as of 2015. In each city, population densities are normalized by the highest 1970 value. The curves in the panels indicate Gaussian kernel densities. (M) The log ratio of each city's population share, between 2015 and 1970, against the log population share to the national population in 2015, with a fitted line by an ordinary least squared regression. (N) The 2015/1970 ratio of the spacing between the r largest cities against the city's population ranking in 2015. The dashed curve indicates the average random counterfactual ratios. The lower (upper) dash-dot curve indicates the 1\% (99\%) value of the random counterfactual ratio.
 The four largest cities $(r \le 4)$ are omitted, as they remain the same between 1970 and 2015.
 \label{fig:size_evolution}}
\end{figure*}

Figs.~\ref{fig:size_evolution}A--D show the 2015/1970 ratios for population size, area, highest population density, and average population density, respectively, for the 302 cities that existed in both 1970 and 2015. 
Population size increased by 21\% on average (Fig.~\ref{fig:size_evolution}A), indicating that the population concentration in the 302 cities increased during the analyzed period. 
Simultaneously, the city area almost doubled, peak population density halved (Figs.~\ref{fig:size_evolution}B and C), and average population density decreased by 35\% (Fig.~\ref{fig:size_evolution}D), indicating the apparent tendency of local dispersion or suburbanization within each city.  
Figs.~\ref{fig:size_evolution}E--L compare the population density distributions between 1970 and 2015 in each of the largest eight cities as of 2015. 
The density distributions shifted to the left, indicating that the local dispersion took place over a larger area in these cities, thus, decreasing the peak population density.

Fig.~\ref{fig:size_evolution}A also suggests that, among the 302 cities, some attracted more population, and others less so.  
Fig.~\ref{fig:size_evolution}M shows the population growth rates for the cities in Fig.~\ref{fig:size_evolution}A against their population shares for 2015. 
There is a trend for larger cities attract more residents, while the opposite holds for smaller cities. 
Fig.~\ref{fig:size_evolution}N reports the tendency of increasing spatial separation for the top 100 cities (see the \textit{SI Appendix} for details). 
Specifically, it shows the ratio $d_r\equiv d_{r,2015}/d_{r,1970}$, where $d_{r,t}$ is an index of the distance between the $r$ largest cities in year $t$, together with the significance levels of their magnitudes. 
If $d_r > 1$, the $r$ largest cities in Japan were spatially more distant from each other in 2015 than they were in 1970. 
We observe that $d_r$ is consistently greater than unity, with $d_r$ being larger for smaller $r$ values. 
Generally, cities became more distant from each other during 1970--2015, with this tendency being stronger for larger cities.

\section*{Discussion}

We identified contrasting evolutions of the urban population in Japan over the study period. 
First, fewer cities grew larger, and they became more spatially distant from each other. 
Second, the remaining cities on average flattened out towards extinction. 

In extant economic models, spatial patterns are explained through the interactions between agents'  positive and negative feedback, mediated by the transport costs between locations. 
Three representative feedback mechanisms can be found in the literature: 
(i) local positive agglomeration effects (i.e., short-range positive feedback) such as knowledge exchanges, market sharing, and production \cite{Duranton-Puga-HB2004}; 
(ii) crowding effects (i.e., short-range negative feedback), such as traffic congestion and local scarcity of land for housing and production, which promote the flattening or suburbanization of a city \cite{Fujita-Book1989}; 
and (iii) competition between different cities (i.e., long-range negative feedback) \cite{%
Krugman-EER1993}. 
In a model with all three effects, lower transport costs promote long-range negative feedback by increasing the competition between more distant locations; 
therefore, only the larger and more distant cities enjoying more extensive local positive feedback, such as Tokyo and Osaka, can thrive.
At the same time, better transport accessibility induces suburbanization within each city as it mitigates the crowding effects \cite{Akamatsu-et-al-DP2020}.  
Overall, the evolution of the urban system in Japan is consistent with these predictions: Japanese cities move towards a mono-centric distribution around Tokyo, which is itself experiencing an evening out, partly confirming the ``death of distance'' \cite{Cairncross-Book1997}.

Since cities can be seen as peaks (spots) in the population distribution, our findings can be associated with the mathematical pattern formation theory \cite{Murray-Book1989}. 
Several analogies exist between our results and known facts in the literature. 
Namely, stationary multiple spots in reaction--diffusion systems arise from a combination of long-range negative and short-range positive feedback \cite{Meinhardt-Gierer-Bioessays2000}. 
When the diffusion coefficients in a reaction-diffusion system become large, the number of spots declines due to ``overcrowding instability'' that limits the possible number of spots in the system \cite{Wei-Winter-JMB2008}. 
Economic models of cities are formulated on the basis of transport costs between locations, which would be interpreted as the impedance for the ``diffusion'' of positive and negative feedback effects. 

Despite the similarities, however, pattern formation in the city system poses new challenges that call for the synthesis of knowledge in economics, biology, and mathematics. 
City sizes exhibit significant variation; their distribution can be approximated well by the power law within many countries \cite{Gabaix-Ioannides-HB2004}. 
This fact invalidates the na\"{i}ve interpretation of cities as multiple-spot patterns in two-component reaction--diffusion models, as no significant variations in spot sizes can arise. 
Economic modeling with homogeneous agents also cannot explain cross-sectional size variations of the cities by self-organization \cite{Akamatsu-et-al-DP2020}. 

An economic rationale for size diversity of cities lies in industrial diversity, which gives rise to the diversity in the spatial extents of feedback effects. For example, how geographically far a firm could locate while still serving customers' needs is different across industries, which leads to the characteristic ``frequency'' in the location of firms in each industry. 
Additionally, many industries coordinate spatially in cities as they seek short-range positive feedback (e.g., cities' large labor and consumer pools). 
This process leads to the simultaneous emergence of spatial fractal structure (or the urban hierarchy) and a power-law city size distribution \cite{Hsu-2012}; both are confirmed in the real-world data \cite{Mori-et-al-PNAS2020}. 
Thus, to explain the observed variation in city sizes and its long-run dynamics, inter-temporal changes in transport costs alone are not sufficient; we should model (industrial) diversity in the spatial extent of feedback effects. 
Research on reaction--diffusion models with more than two components is relatively scarce in the literature owing to their intractability. 
However, in light of the discussed analogies, it would be possible to %
build a unified theory for the city system and other socioeconomic phenomena.

\matmethods{%

\subsection*{Cities}
The Grid Square Statistics of the Population Censuses of Japan provides population count data for 1970--2015 in 30''-by-45'' ($\approx$ 1km-by-1km) grid cells. 
We algorithmically identify a \emph{city} as a set of contiguous cells, each with a density of at least 1,000 people/km$^2$, that yields a total population of at least 10,000. 
See the \textit{SI Appendix} for the data and codes. 

\subsection*{Distance between cities}
The measurement of distance between the $r$ largest cities used in Fig.\ref{fig:size_evolution}N is computed with the Open Source Routing Machine (OSRM, \url{http://project-osrm.org/}) and OpenStreetMap (\url{http://download.geofabrik.de/}). 
See the \textit{SI Appendix} for the formula for $d_{r,t}$ and other details.  

\subsection*{Data availability} 
All data and codes are provided in the \textit{SI Appendix}. 

}
\showmatmethods{} %
\acknow{This research was conducted as part of the project, ``Agglomeration-based framework for empirical and policy analyses of regional economies,'' undertaken at the Research Institute of Economy, Trade and Industry. 
This research has also been supported by the Kajima Foundation, and the Murata Science Foundation, the International Joint Research Center of Advanced Economic Theory of the Institute of Economic Research in Japan, and  the Grant in Aid for Research Nos.~17H00987 and 19K15108 of the MEXT, Japan. Part of this research was conducted under the project, ``Research on the evaluation of spatial economic impacts of building bus termini'' (Principal Investigator: Prof. Yuki Takayama, Kanazawa University), supported by the Committee on Advanced Road Technology, the MLIT, Japan.
}
\showacknow{} %
\bibliography{SpatialScale}

\section*{Supplementary Information} 
\subsection*{Distance between cities}
The distance between cities is computed as the shortest-path road distance between the most densely populated grids within each city. 
We computed bilateral road distances by applying an open-source routing engine, the Open Source Routing Machine (OSRM, \url{http://project-osrm.org/}) to the geographic data of OpenStreetMap (\url{http://download.geofabrik.de/}).
Specifically, we used the routing service version 1 of OSRM with driving mode; the other settings of routing were taken from the OSRM default as described in Supplementary Information of \cite{Mori-et-al-PNAS2020}. The complete manual for the distance computation is available from \textsf{distance\_computation.zip} at \url{https://datadryad.org/stash/dataset/doi:10.5061/dryad.8gtht76k5}.

\subsection*{A measure of distance between the largest cities}
We define a measure of distance between the $r$ largest cities in year $t$ by
\begin{equation}
    d_{r,t} \equiv \frac{1}{r}\sum_{i\in U_{r,t}} \min_{j\in U_{r,t}\backslash \{i\}} d(i,j),\label{eq:spacing}
\end{equation}
where $U_{r,t}$ is the set of the $r$ largest cities in year $t$, and $d(i,j)$ be the road-network distance between cities $i$ and $j$, where the location of a city is the grid cell with the highest population density. 
To gauge the significance of the magnitude of $d_r$, we consider a hypothetical value $\tilde{d}_r$ obtained from \eqref{eq:spacing} with $U_{r,t}$ replaced by $r$ cities randomly selected from all the 886 cities existed in at least one of every five years between 1970 and 2015.

\subsection*{city\_data.zip} 
The data set and Python programs for replicating the results are available from \url{https://www.dropbox.com/s/11ubodqntsu1fvn/city_data.zip?dl=0}.
There are five comma-separated value files:
(i) \textsf{city\_population.csv} contains four column data. Column  \textsf{CITY} includes the city indices, \textsf{POP} the population size, \textsf{NORM\_POP} the population share in the national total population, \textsf{YEAR} the year (1970 or 2015).
(ii) \textsf{balanced\_city\_set.csv} contains the list of cities that existed in both 1970 and 2015. 
(iii) \textsf{cells\_in\_cities.csv} contains four column data of the 1 km-by-1 km grid level population size in each city that existed both in 1970 and 2015. Column \textsf{CELL\_ID} includes unique indices for 1 km-by-1 km grids, \textsf{CITY} the city indices, \textsf{POP} the population sizes of cities, \textsf{NORM\_POP} the shares of cities in the national total population, \textsf{YEAR} the year (1970 or 2015). 
(iv) \textsf{all\_cities\_1970-2015.csv} contains the indices of the most populated 1 km-by-1 km grid cells in the cities that existed any year between 1970 and 2015 (every five years).
(v) \textsf{bilateral\_distances.csv} contains bilateral distances among the grid cells listed in (iv).

There are three Python programs (Python version 3.8). (a) \textsf{CitySizeEvolution.py} generates Fig.~2A-D. (b) \textsf{CitySpatialEvolution.py} generates Fig.~2E-L. (c) \textsf{GlobalConcentration.py} generates Fig.~2M-N. 
Put all programs and the data in the same folder.

\end{document}